\documentclass[runningheads]{llncs}
\usepackage{graphicx}
\usepackage{multirow}
\usepackage{balance} 
\usepackage{stfloats}
\usepackage{xcolor}
\usepackage{amsmath}
\usepackage{amssymb}

\usepackage{caption}
\usepackage{enumitem}
\usepackage{xspace}
\usepackage{fancybox}

\usepackage{url}
\usepackage{listings}
\usepackage{mdframed}
\usepackage{hyperref}
\usepackage{makecell}

\usepackage[T1]{fontenc}
\usepackage{graphicx}
\begin{document}
\title{Database-Augmented RAG \\
for Automated Repair of REST API Misuses\\}
%
%


\author{Shoei Inoue\inst{1} \and
Norihiro Yoshida\inst{1}\orcidID{0000-0003-4910-1729} \and \\
Erina Makihara\inst{1}\orcidID{0009-0008-8777-619X} \and
Shiyu Yang\inst{1}\orcidID{0009-0005-0955-3431} \and
Katsuro Inoue\inst{1}\orcidID{0000-0001-5424-0614}}
\authorrunning{Inoue et al.}
%
\institute{Ritsumeikan University, Osaka, Japan\\
\email{is0604fp@ed.ritsumei.ac.jp}, \email{norihiro@fc.ritsumei.ac.jp},\\
\email{makihara@fc.ritsumei.ac.jp}, \email{yangsy@fc.ritsumei.ac.jp}, \email{inoue-k@fc.ritsumei.ac.jp}}
\maketitle

\begin{abstract}
Many Internet of Things (IoT) services provide Representational State Transfer (REST) APIs, which require client developers to implement applications that conform to the corresponding API specifications.
When client programs contain API misuse, developers debug them based on error responses.
However, such responses are often insufficient for identifying the root cause, requiring developers to repeatedly communicate with the server.
Retrieval-Augmented Generation (RAG) is a promising approach for providing large language models (LLMs) with external knowledge.
However, in automated repair of REST API misuses, it remains unclear how specifications should be stored in a RAG database.
This study evaluates how different configurations for organizing API specifications affect RAG-based repair of REST API misuse.
We constructed 11 RAG configurations with different database structures and compared their repair rates with a baseline method.
For evaluation, we used REST API misuse cases collected from real-world repositories.
The results show that, in the studied datasets, the baseline method achieved a repair rate of 54.3\%, whereas a RAG-based method using four databases achieved a maximum repair rate of 88.6\%.
These results indicate that organizing specifications according to version and content type can be an effective design choice for RAG-based REST API misuse repair.

\end{abstract}

\section{Introduction}
In recent years, Internet of Things (IoT) devices have become increasingly widespread \cite{makhshari2021iot} \cite{iot2}. APIs play a fundamental role in enabling interoperability and scalability, as they rely on continuous communication between heterogeneous devices and cloud or edge services \cite{makhshari2021iot}. Consequently, API-based development approaches have attracted growing attention.

Client development for IoT devices requires consideration of factors such as limited computational resources, diverse network protocols, and heterogeneous entity integration \cite{makhshari2021iot}.
Consequently, IoT systems are more complex than traditional systems, rendering them more prone to errors during development \cite{jiang2020experimental}.
In practice, client implementations must bridge device-specific capabilities, unstable edge networks, and frequent vendor-side specification changes \cite{LERCHER2024112110} \cite{yasmin2020first}.
A considerable portion of these errors remains unresolved, negatively affecting system reliability and maintainability.

Most IoT services provide Representational State Transfer (REST) APIs \cite{fielding2000architectural}, enabling developers to access resources efficiently.

Developers must follow the REST API specifications when implementing client code.
Generally, they send HTTP requests and examine the corresponding responses to verify code correctness.
However, error messages and status codes provide limited diagnostic information and do not indicate the root cause of a failure.
Consequently, developers must repeatedly send requests and check responses to identify and resolve defects in client code.
This ambiguity creates trial-and-error debugging cycles, which increase development time and raise the risk of shipping incomplete fixes.

When providers update a REST API specification, developers must update their client code accordingly.
However, in practice, such updates are often neglected, potentially introducing various defects and vulnerabilities \cite{bodenheim2014evaluation} \cite{jiang2020experimental} \cite{LERCHER2024112110} \cite{yasmin2020first}.
Therefore, developers must promptly update their code to align with the latest REST API specification.

Previous studies explored API misuse detection for Java APIs \cite{sven2019investigating} \cite{li2024boosting} \cite{ren2020api}, as well as automated bug fixing using large language models (LLMs) \cite{ding2024cycle} \cite{kechagia2021evaluating} \cite{xia2023automated} \cite{yin2024thinkrepair}.
These studies focused on method-call misuse in Java.
However, REST API calls differ significantly from Java API calls.
Java APIs need correct method calls. In contrast, REST APIs need correct endpoints, request headers, and request bodies.
Moreover, REST API misuse is often associated with provider-side policy changes, such as authentication updates and version-specific revisions.
Therefore, effective repair requires up-to-date external specification context, rather than relying only on local code semantics.

To address this problem, we adopt Retrieval-Augmented Generation (RAG) \cite{lewis2020retrieval}, which can add domain-specific information and repair details to the prompt, rendering it a promising approach to mitigate these challenges \cite{xu2025rag}.
In this study, we propose a Database-Augmented RAG approach to the automated repair of REST API client programs. We compare its performance with that of the baseline method.

We collected misuse cases from \textit{SwitchBot} and \textit{Fitbit}, both of which provide REST APIs for IoT services. Subsequently, we applied our method to these cases and evaluated 11 RAG database configurations.
Specifically, we designed various data storage strategies, such as storing the latest and deprecated specifications in separate databases, as well as separating code snippets and natural-language text.

The contributions of this study are as follows:
\begin{itemize}
    \item We evaluated 11 distinct database configurations for RAG and compared their effectiveness for repairing REST API misuse in the studied dataset.
    \item We demonstrate that separating retrieval contexts by API version and content type can improve repair performance in the studied dataset.
\end{itemize}

The remainder of this paper is organized as follows.
Section 2 describes key terminology and provides background information.
Section 3 explains the investigation method, dataset, and evaluation metrics.
Section 4 presents the experimental results.
Section 5 discusses related work, and Section 6 concludes the paper with a summary and future directions.

\section{Background}

\subsection{REST APIs}

Numerous IoT device development efforts adopt web APIs based on the REST architectural style \cite{fielding2000architectural}.
This enables developers to access IoT device resources easily.

Representative examples of REST APIs used in IoT services include those provided by \textit{SwitchBot} and \textit{Fitbit} APIs.

The \textit{SwitchBot} API \cite{switchbot} enables control of IoT devices such as smart locks and lighting.
The API underwent version changes in 2022, and the provider deprecated specifications from older versions for newer products.
Table \ref{tab2} lists part of the REST API specification.
The entries in regular font are parameters introduced in version 1.0, whereas those in bold were added in version 1.1.
In this update, the provider introduced new request header parameters, i.e., \textit{sign}, \textit{t}, and \textit{nonce}.

The \textit{Fitbit} API \cite{fitbit} enables users to record and retrieve fitness information through wearable devices.
Although \textit{Fitbit} does not disclose precise version timelines, several features have been deprecated owing to the transition from OAuth 1.0 to OAuth 2.0 and modifications in version 1.2.
Inconsistencies between client code and these specifications may result in REST API misuse.

The \textit{SwitchBot} specifications are publicly available on GitHub \cite{switchbot}, while \textit{Fitbit} publishes its specifications through its official developer reference \cite{fitbit}.

\begin{table}[tb]
\caption{The \textit{SwitchBot} API specification\protect\footnotemark{}}
\label{tab2}
\centering
\begin{tabular}{l|rrr}\hline\hline
Parameter      &  Type           & Location          & Required     \\\hline
Authorization  & String          & header            & Yes          \\
\textbf{sign}  & \textbf{String} & \textbf{header}   & \textbf{Yes} \\
\textbf{t}     & \textbf{Long}   & \textbf{header}   & \textbf{Yes} \\
\textbf{nonce} & \textbf{Long}   & \textbf{header}   & \textbf{Yes} \\\hline
\end{tabular}
\end{table}
\footnotetext{The bold text indicates parts of the REST API specification added in version 1.1.}

\begin{figure}[htbp]
\centering
\begin{lstlisting}[
    language=Python,
    basicstyle=\ttfamily\scriptsize,
    frame=lines,
    xleftmargin=1em,
    breaklines=true,
    lineskip=2pt,
    belowskip=3.5em
]
- BASE_END_POINT = 'https://api.switch-bot.com/v1.0'
+ BASE_END_POINT = 'https://api.switch-bot.com/v1.1'

+ def headers() -> dict:
+     {...}
+     return {
+         'Authorization': token,
+         't': str(t),
+         'sign': str(sign, 'utf-8'),
+         'nonce': nonce
+     }

  devices = requests.get(
      url=BASE_END_POINT + '/devices',
-     headers={
-         'Authorization': os.environ['SWITCH_BOT_OPEN_TOKEN']
-     }
+     headers=headers()
  ).json()['body']
\end{lstlisting}
\caption{Example of \textit{SwitchBot} API misuse and its correction\protect\footnotemark}
\label{switch_repair}
\end{figure}

Figure \ref{switch_repair} shows an example of a misuse resulting from the use of a deprecated version of the \textit{SwitchBot} API specifications, along with its corrected version.
Lines beginning with a minus sign represent the incorrect code, whereas those starting with a plus sign indicate the corrected code.
The misuse occurs at the endpoint specified on line 1 and in the request headers shown in lines 15--17.
In addition to the Authorization header, the updated specification requires \textit{sign}, \textit{t}, and \textit{nonce}.
The corrected version updates the endpoint on line 2 and introduces a function that returns the complete set of required request headers in lines 4--11.
This example demonstrates that developers must update both endpoints and the request headers for the latest API version.

\footnotetext{\href{https://github.com/snoozers/lazy-home/commit/bb9a929d723cc51b67cf08a09de2d183406b3569}{https://github.com/snoozers/lazy-home/commit/bb9a929d723cc51b67cf08a09de2d1
83406b3569}}

\subsection{Existing LLM-based Automated Program Repair Methods and Their Limitations}
In existing research, Xia et al. \cite{xia2023automated} compared traditional automated program repair tools, such as TBAR \cite{liu2019tbar} with LLM-based repair techniques.
They evaluated nine pretrained LLMs on five datasets across three programming languages and showed that LLMs outperform traditional repair tools in terms of language generality and repair accuracy.
However, repairs that require external information, such as REST API specifications, fall outside their scope.

Li et al. \cite{li2024exploring} proposed a parameter-efficient fine-tuning (PEFT) method that updates only a subset of model parameters. They aimed to evaluate the effectiveness and efficiency of PEFT relative to full-model fine-tuning (FMFT). Using three benchmark datasets, they showed that PEFT can maintain or improve repair performance while using fewer computational resources.

The two aforementioned approaches differ in how they balance external knowledge, maintenance cost, and adaptability. A standalone LLM is easy to set up but lacks access to up-to-date external knowledge. Fine-tuning improves domain knowledge; however, it incurs substantial maintenance costs.

\subsection{RAG}
We build on existing automated program repair methods by incorporating RAG.
RAG retrieves relevant text segments and injects them into prompts, enabling LLMs to use information that is not fully captured in pretraining data \cite{xu2025rag}.

Conventional RAG typically stores all documents in a single database.
However, this design risks confounding the deprecated and latest specifications or intermixing code snippets with natural-language text.
For repairing REST API misuse, this can reduce the repair rate because the model may receive context that is partially correct but temporally inconsistent.
When deprecated and latest specifications are embedded together, semantically similar but deprecated segments may dominate retrieval results, reducing prompt reliability even if retrieval scores are high.

Therefore, we propose a Database-Augmented RAG approach that partitions specification content into multiple databases by version and information type.
This design aims to improve retrieval precision and provide more controllable prompt context for repair generation.

In this study, we evaluated how database configuration affects repair rates and identified the most effective configuration for correcting REST API misuse.

\begin{table*}[t]
    \centering
    \caption{Configurations for storing REST API specifications in databases}
    \label{pattern}
    \small
    \setlength{\tabcolsep}{3.5pt}
    \begin{tabular}{l|cccccc|c}
        \hline\hline
        & Deprecated & Latest
        & \makecell{Deprecated\\code}
        & \makecell{Deprecated\\text}
        & \makecell{Latest\\code}
        & \makecell{Latest\\text}
        & \makecell{Prompt\\Types} \\
        \hline
        \textbf{BL} & & & & & & & p1 \\
        \hline
        \textbf{D} & \checkmark & & & & & & p2 \\
        \hline
        \textbf{L} & & \checkmark & & & & & p2 \\
        \hline
        \textbf{DLm}\protect\footnotemark{} & \checkmark & \checkmark & & & & & p2 \\
        \hline
        \textbf{DL} & \checkmark & \checkmark & & & & & p2 \\
        \hline
        \textbf{Ds} & & & \checkmark & \checkmark & & & p2+p3 \\
        \hline
        \textbf{Ls} & & & & & \checkmark & \checkmark & p2+p3 \\
        \hline
        \textbf{DLs} & \checkmark & & & & \checkmark & \checkmark & p2+p3 \\
        \hline
        \textbf{DsL} & & \checkmark & \checkmark & \checkmark & & & p2+p3 \\
        \hline
        \textbf{DsLs} & & & \checkmark & \checkmark & \checkmark & \checkmark & p2+p3 \\
        \hline
        \textbf{DcLc} & & & \checkmark & & \checkmark & & p2+p3 \\
        \hline
        \textbf{DtLt} & & & & \checkmark & & \checkmark & p2+p3 \\
        \hline
    \end{tabular}
\end{table*}

\footnotetext{This shows a single database configuration that stores all specifications.}

\section{Proposed Method}
Recent studies have applied various RAG configurations to automated program repair \cite{fan2023automated} \cite{jin2023inferfix} \cite{xia2023automated}. However, they have not examined which database configuration is effective for automatically correcting REST API misuse. Therefore, this study evaluates 11 database configurations to identify an effective RAG architecture.

Table \ref{pattern} presents the 11 configurations. The naming conventions are defined as follows:

\begin{itemize}
\item \textbf{BL (Baseline method)}: The baseline method performs correction without RAG.
\item \textbf{Specification Type}: \textbf{D} denotes deprecated specifications, and \textbf{L} denotes latest specifications.
\item \textbf{Storage Strategy}: \textbf{m} denotes mixed storage of D and L in a single database, \textbf{s} denotes separate storage of code snippets and natural-language text, \textbf{c} denotes code-only storage, and \textbf{t} denotes text-only storage.
\end{itemize}

A configuration name without storage strategy symbols, such as \textbf{D} or \textbf{L}, indicates that the specifications are stored in a single database. For example, \textbf{D} stores only deprecated specifications in one database. In contrast, \textbf{DLs} stores deprecated specifications in one database and separates the latest specifications into code snippets and natural-language text using three databases.

Based on Table \ref{pattern}, we first describe the baseline method (\textbf{BL}), followed by the single-database RAG method (\textbf{DLm}) and the multi-database RAG method (\textbf{DsLs}). Details of the remaining configurations are provided in our GitHub repository\footnote{\url{https://github.com/shonsukee/database-augmented-rag}}.

All 11 configurations use prompt templates with basic instructions and the target program. Since the correction procedure differs by database configuration, we customize the prompt for each method. As shown in the ``Prompt Types'' column of Table \ref{pattern}, the prompts use three types of supplementary information: \textbf{p1} inserts a link to the REST API specification, \textbf{p2} inserts relevant parts of the specification, and \textbf{p3} separately inserts code snippets and natural-language text.

\subsection{Baseline Method}
\textbf{BL} is used as the baseline for evaluating the effect of RAG-based specification retrieval.
It uses a prompt consisting of basic instructions, the target program, and links to the REST API specifications. Because the full API specifications cannot fit within GPT-4o's context window, we provide specification links rather than inserting the entire specification contents into the prompt. GPT-4o can then access the linked specifications through its built-in web search capability during repair generation. This prompt is directly input into GPT-4o without RAG.

\begin{figure*}[t]
    \centering
    \includegraphics[scale=0.36]{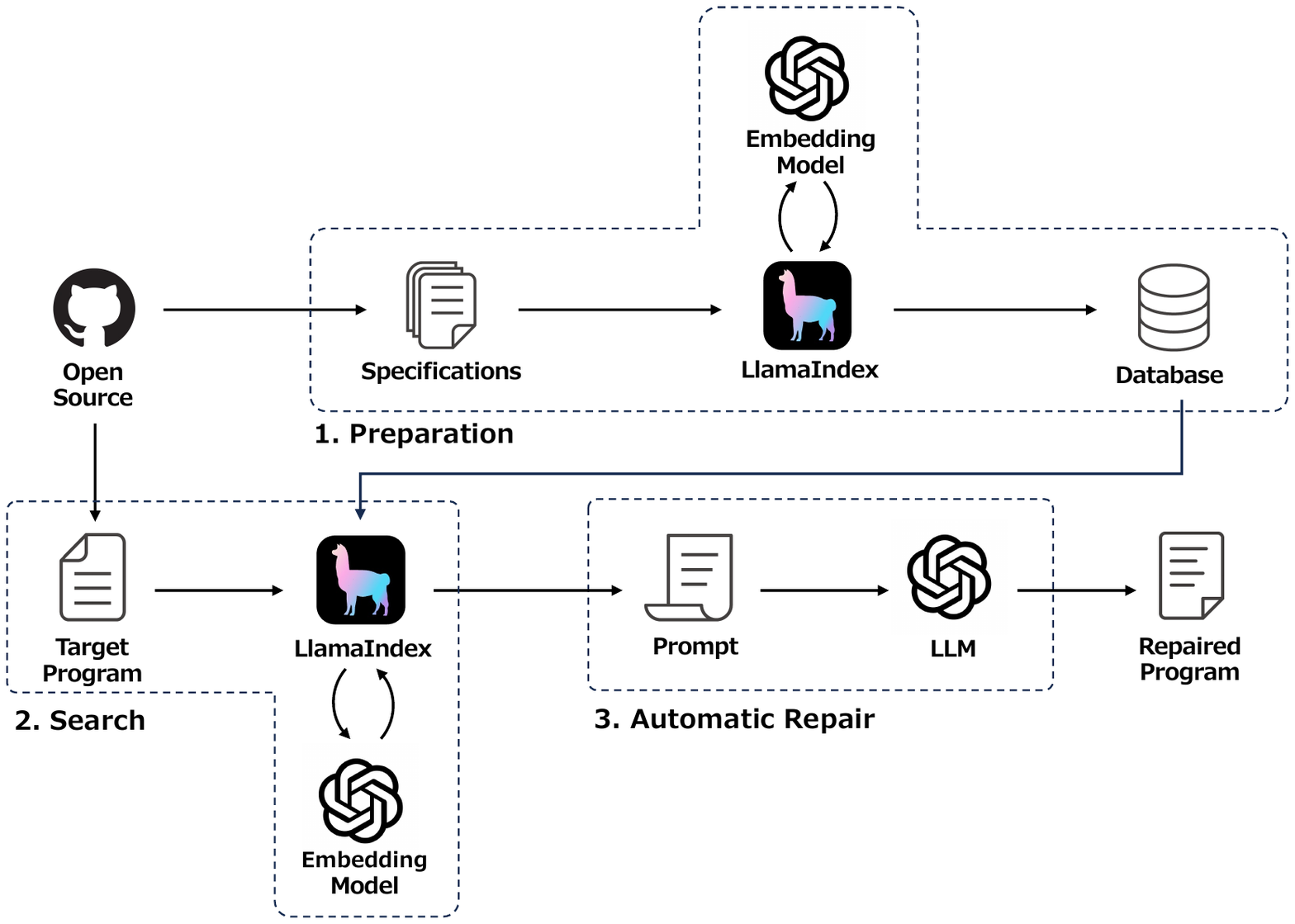}
    \vspace{30pt}
    \caption{Overview of the Standard RAG Method}
    \label{standard}
\end{figure*}

\subsection{Standard RAG Method}
The baseline method has several limitations, such as incorrect link selection and the retrieval of irrelevant information, which can reduce the repair rate.

To address these limitations, we use the RAG approach in the \textbf{DLm} configuration.
Figure \ref{standard} shows an overview of this standard RAG method.

First, we collect REST API specifications from open-source platforms such as GitHub and extract both natural-language text and code snippets. The extracted data are then split into chunks using LlamaIndex, a framework for data indexing and retrieval.
In this process, the specification text is initially segmented at the paragraph level. When a segment exceeds the chunk size, the method further subdivides it using sentence boundaries, regular expressions, whitespace, and, finally, character-level splitting. We set the chunk size to 2048 tokens and the overlap length to 200 tokens to preserve related specifications within a single chunk. Finally, the chunks are vectorized using OpenAI's text-embedding-3-large model and stored in a vector database.
This process enables semantic search over the specifications, which is beneficial because client code and API specifications may use different terminology for the same operations.

Next, we collect target functions containing REST API misuse instances for \textit{SwitchBot} and \textit{Fitbit} using the GitHub API.
These target functions are embedded using the same embedding model used for the specification chunks.
We then perform a semantic search against the database using the vectorized target functions and retrieve the top 60 relevant segments.
When limited relevant context is available, fewer than 60 segments may be retrieved.

Finally, we construct a prompt for the LLM using the retrieved text segments and the target program.
The prompt is designed based on the prompt engineering guide \cite{prompt}, and further details are provided in our GitHub repository\footnote{\url{https://github.com/shonsukee/database-augmented-rag}}.
The prompt includes background information, repair instructions, database information, and the target program.
We also adjust the prompt structure based on the retrieved information type, such as deprecated or latest specifications, and define the output format.
This prompt is then input into GPT-4o to generate the final correction.

\begin{figure*}[t]
    \centering
    \includegraphics[scale=0.36]{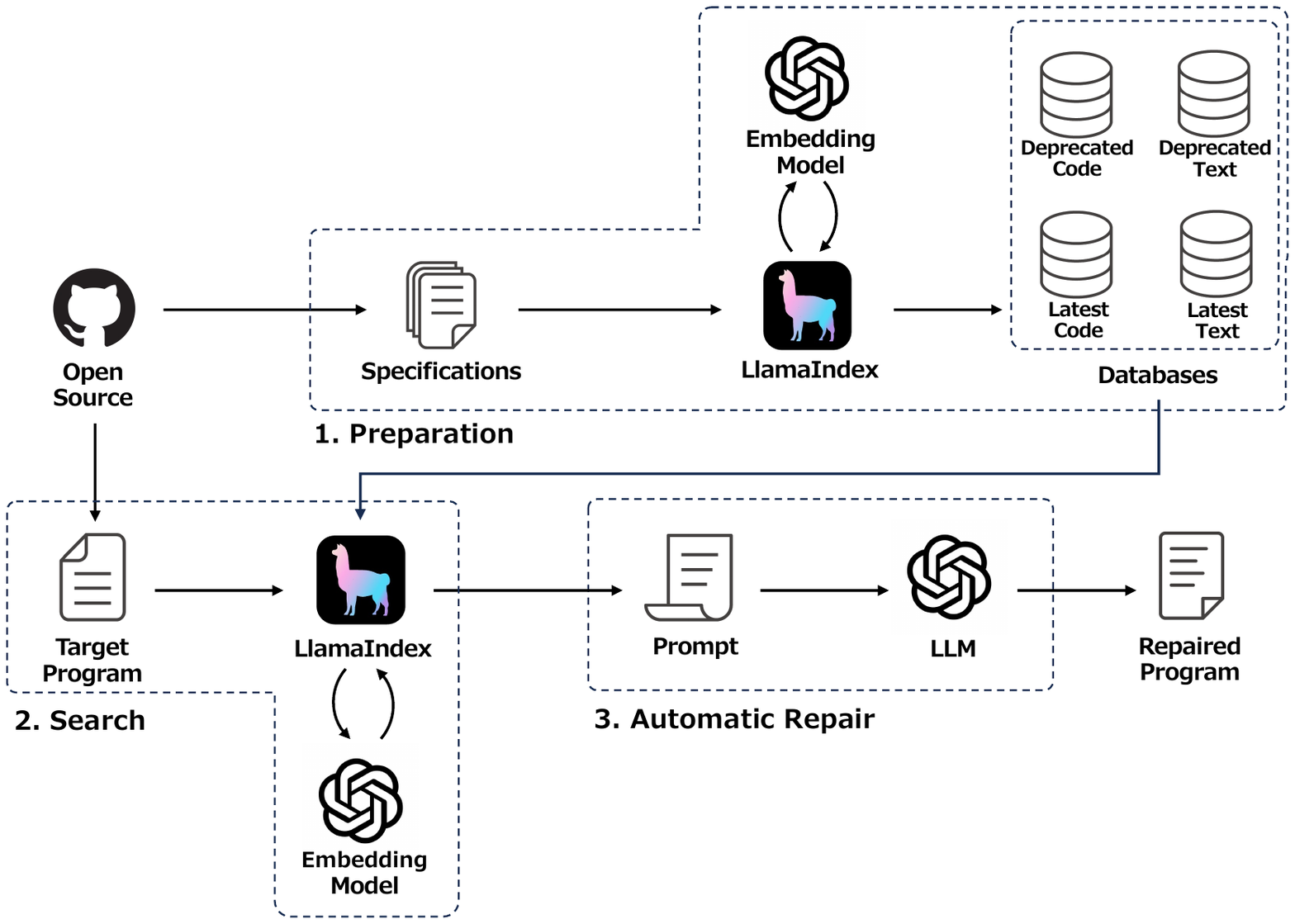}
    \vspace{30pt}
    \caption{Overview of the Database-Augmented RAG Method}
    \label{fig4}
\end{figure*}

\subsection{Database-Augmented RAG Method}
The standard RAG method retrieves semantically relevant text segments; however, Figure \ref{standard} shows that deprecated and latest specifications are stored in a single database. This design combines the latest and deprecated specifications, which makes it difficult to retrieve all necessary information for the correction.

To improve retrieval precision, we partition the specification content into four databases: code snippets from deprecated specifications, natural-language text from deprecated specifications, code snippets from latest specifications, and natural-language text from latest specifications. Figure \ref{fig4} shows the \textbf{DsLs} configuration, which extends the standard RAG method by partitioning information according to version and content type.

This method differs from the standard RAG method in database storage, text extraction, and prompt design.

First, we separate the REST API specifications into code snippets and natural-language text. We extract code snippets using HTML tags, while natural-language text is obtained by removing code snippets from the specifications. We then embed these data using OpenAI's text-embedding-3-large model and store them in the corresponding databases.

Next, we search each of the four databases and retrieve the top 60 segments from each database. We use the same chunking and top-k settings as the standard RAG method for all four databases. Finally, we adjust the prompt structure for this configuration, as shown in Table \ref{pattern}. The prompt consists of three stages. In Stage 1, the model analyzes deprecated code snippets and natural-language text. In Stage 2, it compares the target program with the deprecated specifications to identify misuse instances. In Stage 3, it uses the latest code snippets and natural-language text to generate the corrected code. This staged prompt design helps guide the model and reduce reasoning errors.

\subsection{Construction of Evaluation Dataset}
To construct the dataset, we collect source code from GitHub repositories that use the \textit{SwitchBot} and \textit{Fitbit} APIs. We also collect closed commits, issues, and pull requests related to these APIs using the GitHub API.

Specifically, we submit search queries to the GitHub API to collect links to closed commits, issues, and pull requests. The queries include terms such as ``switchbot api update,'' API domain names, and keywords related to deprecated APIs. We then filter the collected links using automated programs and manual inspection.

In this filtering process, we excluded cases that were not suitable for specification-based REST API repair. Specifically, we removed syntax errors, project-specific cases, and cases whose correctness could not be determined from the official specifications. Next, the first author selected commits and pull requests that update code to the latest REST API specifications, as well as issues that reference deprecated specifications or cite closed commits. Finally, the first author manually extracted programs using deprecated specifications. During this process, the first author verified whether the changes directly correspond to specification updates. Consequently, we obtained nine cases involving both request header and endpoint misuse, and twelve cases involving endpoint-only misuse.

\subsection{Evaluation}
The success rate was calculated as the ratio of successful patch generations to the total number of execution runs. Specifically, for each case, we counted the number of successful patches across five executions and divided it by the total number of executions.

A case is considered successful if it satisfies the rules defined in existing research \cite{liu2020efficiency} and matches the corresponding GitHub repository\footnote{\url{https://github.com/TruX-DTF/APR-Efficiency}}. We manually compared the repository code, latest specification, and generated patch. We applied the LLM-generated code to the original repository and checked whether the repository compiled successfully. A successful patch must fix all misuse instances in the target function.

This paper presents only the results of each method due to space limitations. Further details are available in our GitHub repository\footnote{\url{https://github.com/shonsukee/database-augmented-rag}}.

To identify the highly effective RAG configuration, we formulate the following research question:
\begin{description}
    \item [RQ:] {\bf Which RAG database configuration is highly effective?}\\
    We varied the number and type of databases in the RAG architecture to enhance the repair rate.
\end{description}

\begin{table*}[t]
    \centering
    \caption{Number of misuses in the dataset and successful repairs}
    \label{tab5}
    \small
    \setlength{\tabcolsep}{2.4pt}
    \begin{tabular}{l|cccccccccccc}
        \hline\hline
        Misuses & \textbf{BL} & \textbf{D} & \textbf{L} & \textbf{DLm} & \textbf{DL} & \textbf{Ds} & \textbf{Ls} & \textbf{DLs} & \textbf{DsL} & \textbf{DsLs} & \textbf{DcLc} & \textbf{DtLt} \\ \hline
        RH \& EP (9) & 1.8 & 0.0 & 7.4 & 6.4 & 7.4 & 0.0 & 7.6 & 7.8 & 6.8 & 7.8 & \textbf{8.0} & 7.2 \\ \hline
        EP (12)      & 9.6 & 4.4 & 10.6 & 10.6 & 10.4 & 6.2 & 10.4 & 10.4 & \textbf{11.0} & 10.8 & 10.0 & 10.2 \\ \hline
        Total (21)   & 11.4 & 4.4 & 18.0 & 17.0 & 17.8 & 6.2 & 18.0 & 18.2 & 17.8 & \textbf{18.6} & 18.0 & 17.4 \\ \hline
        Rate (\%)    & 54.3 & 21.0 & 85.7 & 81.0 & 84.8 & 29.5 & 85.7 & 86.7 & 84.8 & \textbf{88.6} & 85.7 & 82.9 \\ \hline
    \end{tabular}
\end{table*}

\section{Results}
We applied the proposed method to datasets containing misuse cases from two REST API services and evaluated the results.
Table \ref{tab5} presents the experimental results, where \textbf{RH} denotes request header errors and \textbf{EP} denotes endpoint errors.
The table reports four evaluation dimensions: mixed misuse involving both request headers and endpoints, single-endpoint misuse, total repairs, and the repair rate.

\subsection{Discussion}

The results indicate three main causes of repair failures.
First, we observed cases in which the model repaired only part of the misuse instances when a target function contained multiple errors, such as both endpoint and request header misuse.
This explains the lower repair rate for mixed misuse cases and suggests that one-pass repair is insufficient for multiple independent corrections.
Second, some outputs failed to detect misuse when the code was syntactically valid but inconsistent with the API specification.
Therefore, the prompt should indicate the specification constraints that are violated before repair generation.
Third, the vector search failed to retrieve sufficient context required for the repair.
In failed outputs, we observed unsupported API usage, such as references to a non-existent \texttt{v2.0} API and misuse of \texttt{Bearer} authentication.
When exact evidence from the specifications is missing, generated repairs may introduce unnecessary edits.
Thus, retrieval should focus on the exact differences between the deprecated and latest specifications, as these differences show the correct changes.
These observations indicate that prompt ambiguity and insufficient specification evidence can result in unsupported repair content.

\textbf{D} and \textbf{Ds} include only the deprecated specification in the prompt and show much lower repair rates than the baseline method.
In particular, Table \ref{tab5} shows that both methods achieve a 0\% repair rate for mixed misuse cases involving request headers and endpoints.
This result is consistent with the prompt context encouraging outdated API usage.
The effect is particularly strong for mixed misuse cases because they require multiple correction decisions.
Although public REST API specifications may be included in pretraining data, the results show that explicitly provided deprecated context can still reduce repair accuracy.

\textbf{DLs} and \textbf{DsLs} achieve high repair rates for multiple misuse instances because they include code snippets from the latest specifications in the prompt.
These snippets help the model identify concrete differences between deprecated and latest API usage, which is particularly useful for mixed misuse cases involving both endpoints and request headers.
Therefore, code snippets should be prioritized when repairing complex REST API misuse.
However, the improvement is limited for datasets with only endpoint misuse instances.
For example, \textbf{DsL} achieves the highest repair rate for single-endpoint misuse instances, but the difference from the baseline is only 1.4 cases.
This result suggests that simple endpoint errors require less external context than mixed misuse cases.

Next, we compare \textbf{DLm} and \textbf{DL}.
Both methods use the deprecated and latest specifications, but \textbf{DL} stores them in separate databases.
\textbf{DL} achieves a 3.8\% higher repair rate than \textbf{DLm} and repairs 1.0 more mixed-misuse cases.
This result indicates that separating specification versions helps present deprecated and latest API usage as distinct contexts.
As both methods use the same information, the improvement is likely due to reduced retrieval noise and more consistent patch generation.

Finally, we compare \textbf{DcLc} and \textbf{DtLt}. \textbf{DcLc} repairs 0.6 more cases than \textbf{DtLt} and achieves the highest repair rate for multiple misuse instances among all configurations.
This result indicates that using code snippets as retrieval targets is more effective than relying on textual descriptions in this setting.
Code snippets closely match the structure of misuse code and show the differences between the deprecated and latest specifications.
In contrast, natural-language text may explain the background changes but may not provide enough evidence for locating misuse points and generating concrete edits.
Therefore, retrieval should prioritize code snippets and use natural-language text as secondary support.

Overall, increasing the number of databases in the RAG architecture improves the repair rate for REST API misuse.

\begin{mdframed}[linewidth=1pt,linecolor=black]{Answer to RQ:}
The repair rate increases as more databases are added.
The four-database configuration achieves the highest repair rate of 88.6\%.
Clarifying the context contributes to this improvement.
\end{mdframed}

\subsection{Threats to Validity}
First, the evaluation covers only two API services, \textbf{SwitchBot} and \textbf{Fitbit}, and 21 misuse cases.
This dataset is sufficient to compare configurations under a consistent protocol; however, it remains small for drawing broad conclusions about REST API repair.
In addition, the misuse cases are concentrated on request headers and endpoints.
Therefore, the observed trends may not generalize to other REST API ecosystems, documentation formats, authentication mechanisms, or history-repair patterns.
The 88.6\% repair rate of \textbf{DsLs} suggests that this configuration can be effective in the studied dataset; however, it should not be interpreted as a direct estimate of repair performance across broader real-world REST API settings.
Because the dataset is small, differences between configurations can correspond to only a few cases.

Second, the configuration comparison includes possible confounding between database separation and the amount of retrieved context.
Configurations that separate deprecated and latest specifications, or text and code snippets, may improve retrieval control.
However, \textbf{DsLs} can retrieve segments from multiple databases and may therefore provide more usable context than less separated configurations.
Thus, the improvement may be associated with separation; however, the current design cannot fully isolate whether the gain originates from database structure, context quantity, or their interaction.

Third, the baseline control has limitations.
The baseline relies on web search and links to official specifications, and therefore its failures cannot be attributed only to the absence of RAG.
Search results, accessible documentation, and prompt context may vary across cases.
A more controlled comparison would require a non-RAG baseline that receives the same specification text, number of chunks, and token budget as the RAG configurations.

Fourth, we evaluate each case five times.
Following the evaluation criteria of existing research \cite{liu2020efficiency}, the first author determines success by manually comparing the LLM-generated patch with repository fixes and official specifications.
This procedure follows a fixed rule and supports reproducibility at the protocol level; however, evaluator-dependent judgment can still affect borderline cases.
Future evaluation should strengthen this process using multiple evaluators, an explicit rubric, and automated checks where possible.

\section{Related Work}
In this section, we review existing research relevant to our study, focusing on API misuse, Automated Program Repair (APR) techniques that utilize LLMs, and RAG.

\subsection{API Misuse}
Approaches that use graph structures and constraints to detect Java API misuse have been proposed.
Ren et al. constructed a Java API constraint knowledge graph to detect misuse \cite{ren2020api}.
Li et al. extracted constraints from client code, API specifications, and library code. They built a constraint graph to improve detection accuracy \cite{li2024boosting}.
Sven et al. developed a tool that detects and ranks misuse using an API usage graph \cite{sven2019investigating}.
These studies focused on structural misuse in programming language APIs and do not address REST-specific misuse or automated repair.

Other studies focused on REST APIs in the context of web services.
Atlidakis et al. developed a fuzzing tool that analyzes cloud API specifications to automatically test services. They applied this tool to verify services such as GitLab, Microsoft Azure, and Office 365 \cite{atlidakis2019restler}.
Huang et al. used program analysis to verify REST API compliance with specifications \cite{huang2024generating}.
While these studies effectively detect REST API errors, they do not address automated repair. In contrast, our study targets REST APIs and focuses on automated repair based on REST API specifications.

\subsection{Automated Program Repair}
APR aims to generate patches for specific software defects.
Conventional APR methods can be classified into three categories: heuristic-based \cite{le2016history} \cite{le2011genprog} \cite{wen2018context}, constraint-based \cite{demarco2014automatic} \cite{le2017s3} \cite{mechtaev2016angelix}, and template-based \cite{ghanbari2019practical} \cite{hua2018sketchfix} \cite{liu2019tbar}.
Template-based APR tools have been widely studied and have attracted considerable attention.
These approaches rely on predefined patterns. However, this reliance limits their ability to fix novel errors or perform complex repairs.

With the rapid advancement of LLMs, LLM-based APR approaches have become increasingly popular.
Kechagia et al. evaluated APR tools using a Java API misuse benchmark \cite{kechagia2021evaluating}. Recently, LLM-based repair methods surpassed these conventional tools \cite{brown2020language} \cite{leon2025gpt}.
Fan et al. demonstrated that LLMs can fix more errors than conventional APR tools \cite{fan2023automated}.
Xia et al. demonstrated that LLMs are effective in zero-shot program repair without fine-tuning. Prompt engineering and RAG enable high-precision repairs without costly retraining \cite{xia2022less}.
Yin et al. combined chain-of-thought (CoT) and few-shot prompting. They successfully fixed Java API misuse using this method \cite{yin2024thinkrepair}.

Despite these advances, LLM-based repair methods still face limitations. LLMs may generate hallucinated or incorrect code. Hybrid methods that combine LLMs with other tools have been proposed to address this issue.
Jin et al. combined static analysis tools with LLMs to improve automated repair \cite{jin2023inferfix}.
Wei et al. combined LLMs with a completion engine to enhance repair capabilities \cite{wei2023copiloting}.
Li et al. proposed GiantRepair, a tool that concretizes LLM-generated patches using static analysis \cite{li2025hybrid}.

REST API misuse involves incorrect endpoints and parameter definitions, which depend on external REST API specifications, rather than internal source code.
However, static analysis tools only analyze internal code. Hence, they cannot guarantee compliance with external specifications.
Therefore, we used RAG to directly access specification information, enabling appropriate repairs without relying on test execution or internal code analysis.

\subsection{Knowledge-Augmented Repair}
RAG has also been applied to APR and bug analysis. It retrieves and incorporates external knowledge to improve LLM performance.
Wang et al. repaired Java API misuse using RAG. They searched for pairs of past errors and fixes \cite{wang2023rap}.
Li et al. proposed KEPT, a method that uses a knowledge graph from project documents and past code to localize errors \cite{li2025knowledge}.

These studies primarily rely on past code history or bug-fix pairs as knowledge sources.
Our study searched REST API specifications using RAG. REST API specifications are one of the most reliable sources for correct API usage. We automate the repair process based on these specifications rather than historical repair data.
Moreover, REST API specifications change frequently, rendering change management crucial \cite{lercher2025autoguard}.
Training data and repair histories may become outdated immediately after an API update.
To address this issue, our method directly references the latest specifications to ensure up-to-date and reliable repairs.

\section{Conclusion}
This study examined how database design in RAG affects LLM-based automated repair of REST API misuse.
We compared the baseline method with 11 RAG configurations that vary in specification type and storage strategy.
The baseline achieved a 54.3\% repair rate, while the best-performing configuration reached 88.6\%.

The results show two consistent trends.
First, configurations that relied only on deprecated specifications, such as \textbf{D} and \textbf{Ds}, performed worse than the baseline.
Second, configurations that incorporated the latest specifications generally improved repair accuracy.
In particular, separating evidence by version and content type further improved the repair of multiple misuses.
These findings suggest that accurate LLM-based API misuse repair requires the latest specifications and a retrieval design that enables the model to leverage them effectively.
In our dataset, the latest code snippets were especially important because they helped the model fix multiple errors within a single function.

This study also highlights several directions for future work.
To improve context segmentation, we plan to design a paired retrieval mechanism that jointly retrieves code snippets and their corresponding explanatory text from the same region of the specification.
To address limitations in dataset scale, we plan to expand the dataset to include additional API services and more diverse misuse types.
A larger and more heterogeneous dataset will allow us to examine whether the observed trends generalize beyond the current two-service setting.
Finally, to reduce evaluator dependence, we plan to improve the reproducibility of the evaluation process.
Specifically, future evaluations will combine explicit decision criteria with automated checks, ensuring that manual reviews remain traceable while reducing subjective variation.

\begin{credits}
\subsubsection{\ackname}This work was supported by JSPS KAKENHI Grant Numbers JP24K02923, JP23K28065. 

\end{credits}

\bibliographystyle{splncs04}
\bibliography{reference}

\end{document}